\documentclass[%
 aip,
 amsmath,amssymb,
 reprint,%
]{revtex4-1}

\usepackage{graphicx, xcolor}
\usepackage{dcolumn}
\usepackage{bm}
\usepackage{algorithmic}
\usepackage{algorithm}
\usepackage{multirow}
\usepackage[utf8]{inputenc}
\usepackage[T1]{fontenc}
\usepackage{mathptmx}
\usepackage{etoolbox}
\usepackage{subfig}
\makeatletter
\def\@email#1#2{%
 \endgroup
 \patchcmd{\titleblock@produce}
  {\frontmatter@RRAPformat}
  {\frontmatter@RRAPformat{\produce@RRAP{*#1\href{mailto:#2}{#2}}}\frontmatter@RRAPformat}
  {}{}
}%
\makeatother
\begin{document}

\preprint{AIP/123-QED}

\title[ ]{
Formal $O(N^3)$ scaling $GW$ calculations by block tensor decomposition for large molecule systems}

\author{Yueyang Zhang}%
\affiliation{The State Key Laboratory of Physical Chemistry of Solid Surfaces, Fujian Provincial Key Laboratory of Theoretical and Computational Chemistry, and College of Chemistry and Chemical Engineering, Xiamen University, Xiamen, Fujian 361005, China}

\author{Wei Wu}
\affiliation{The State Key Laboratory of Physical Chemistry of Solid Surfaces, Fujian Provincial Key Laboratory of Theoretical and Computational Chemistry, and College of Chemistry and Chemical Engineering, Xiamen University, Xiamen, Fujian 361005, China}
 
\author{Peifeng Su*}
\email{supi@xmu.edu.cn}
\affiliation{The State Key Laboratory of Physical Chemistry of Solid Surfaces, Fujian Provincial Key Laboratory of Theoretical and Computational Chemistry, and College of Chemistry and Chemical Engineering, Xiamen University, Xiamen, Fujian 361005, China}

\date{\today}

\begin{abstract}

Within the framework of many-body perturbation theory based on Green's functions, the $GW$ approximation has emerged as a pivotal method for computing quasiparticle energies and excitation spectra. However, its high computational cost and steep scaling present significant challenges for applications to large molecular systems. In this work, we extend the block tensor decomposition (BTD) algorithm, recently developed in our previous work [J. Chem. Phys. 163, 174109 (2025)] for low-rank tensor compression, to enable a formally $O(N^3)$-scaling $GW$ algorithm. By integrating BTD with an imaginary-time $GW$ formalism and introducing a real space screening strategy for the polarizability, we achieve an observed scaling of approximately $O(N^2)$ in test systems. Key parameters of the algorithm are optimized on the S66 dataset using the JADE algorithm, ensuring a balanced compromise between accuracy and efficiency. Our BTD-based random phase approximation also exhibits $O(N^2)$ scaling, and eigenvalue-self-consistent $GW$ calculations become feasible for systems with over 3000 basis functions. This work establishes BTD as an efficient and scalable approach for large-scale $GW$ calculations in molecular systems.
\end{abstract}

\maketitle


\section{Introduction}
Many-body perturbation theory (MBPT) based on Green's functions has become a powerful framework for computing ground and excited states in molecule systems.\cite{mbpt1_, mbpt2_} 
In MBPT, the one-body Green's function $\bm{G}$ is obtained by solving the Dyson equation, with all correlation effects encapsulated in the self-energy $\bm{\Sigma}$, which plays a role analogous to the exchange-correlation potential in density functional theory (DFT). Hedin's equations provide a systematic and rigorous framework for constructing the self-energy and performing MBPT calculations.\cite{hedin} The simplest approximation derived from Hedin's equations is  $GW$,\cite{gw_int1, gw_int2, gw_int3_bGW} which has been successfully applied in condensed-matter physics for predicting and correcting electronic band structures and is now increasingly applied to molecular systems.

In the $GW$ approximation, vertex corrections are neglected by setting the vertex functional to unity. For molecular systems, $GW$ approximation is routinely employed to compute the properties such as HOMO–LUMO gaps and ionization potentials (IPs).\cite{gw4mol_1, gw4mol_2, gw4mol_3, gw4mol_4, gw4mol_5, gw4mol_6} Rather than performing fully self-consistent $GW$ (sc$GW$) calculations, one often starts with the one-shot $G_0W_0$ approach. To improve the non-self-consistent $G_0W_0$ approach, iterative schemes such as eigenvalue self-consistent $GW$ (ev$GW$) and quasiparticle self-consistent $GW$ (qp$GW$) have been proposed.\cite{SGW, qpGW, fsGW, sGW0, evGW}  Within this hierarchy, $GW$ becomes an accurate and efficient approach to charged excitation spectra, starting from mean-field reference states to deliver quantitative electronic structure corrections.\cite{bse_ma, bse_jpcl_2020, MRP_2002} 
A notable success of the BSE@$GW$ approach, the combination BSE with $GW$, is its ability to accurately describe charge-transfer excitations, which remain challenging for time-dependent DFT (TD-DFT).\cite{bse4mol_1, bse4mol_2, bse4mol_3, bse4mol_4} 


Although $GW$-based methods are computationally more efficient than advanced post-Hartree-Fock methods like coupled cluster method (CC),\cite{cc1,CC2,cc3, book, mbpt-cc} they are nevertheless more expensive than KS-DFT, hindering their applications to large systems. Without further numerical approximations, the formal scaling of $GW$ calculations is $O(N^6)$. To reduce the cost, many efficient numerical algorithms have been developed to achieve $O(N^4)$,\cite{RI-RPA, on4_gw_1} $O(N^3)$,\cite{gw4mol_2, on3_gw_1, on3_gw_2, THC-RPA1, on3_gw_6} and even linear-scaling by exploring locality or stochastic averaging.\cite{on_gw_1, on_gw_2} Among them, low-rank algorithms are particularly effective in reducing computational cost by decomposing electron-electron interactions. Based on the forms of electron-repulsion integral (ERI) factorization, lower-rank algorithms can be divided into three types:
\begin{subequations}
    \begin{align}
    &\begin{aligned}
           (\mu\nu|\lambda\sigma) \approx& \sum_M(\mu\nu|\tilde{M})(\tilde{M}|\lambda\sigma)\\
        =& \sum_{MN} (\mu\nu|M)(M|N)^{-1}(N|\lambda\sigma)
        \end{aligned}
    \label{RI-format}\\
        &(\mu\nu|\lambda\sigma) \approx \sum_{g}X_{\mu g}X_{\nu g}V^{g}_{\lambda \sigma} \approx \sum_{g}X_{\lambda g}X_{\sigma g}V^{g}_{\mu\nu}\label{SP}\\
        &\begin{aligned}
            (\mu\nu|\lambda\sigma) \approx& \sum_{KL}X_{\mu K}X_{\nu K}X_{\lambda L}X_{\sigma L}V_{KL}\\
            =&\sum_{KL}X_{\mu K}X_{\nu K}X_{\lambda L}X_{\sigma L}\sum_{M}B_{KM}B_{LM}
        \end{aligned}
    \label{THC-format}
    \end{align}
\end{subequations} 
Here eq.(\ref{RI-format}) corresponds to the resolution of identity (RI) \cite{Weigend2002,Hohenstein2010} and Cholesky decomposition (CD); \cite{cd1,cd2,cd2,cd4,Pedersen2009} eq.(\ref{SP}) represents pseudo-spectral methods \cite{Friesner1, Friesner2, Friesner3, Friesner4, MRCISD-SP, GVB-PS, MP3-SP, CID-SP} while eq.(\ref{THC-format}) shows the tensor hyper-contraction (THC).\cite{Hohenstein_2012,thc1,thc2,isdf1,isdf2,isdf3,isdf4} 

These low-rank algorithms have significantly accelerated electronic structure calculations, reducing the computational cost for both DFT and post-Hartree-Fock methods. In MBPT calculations, the use of RI reduces the time scale of $GW$ from $O(N^6)$ to $O(N^4)$ for $GW$.\cite{RI-RPA} THC further reduces the scaling to formal $O(N^3)$ and has been extended to more advanced MBPT methods beyond $GW$.\cite{gw_sosex,THCGF2} Although THC has been successfully applied to periodic $GW$ calculations,\cite{isdf-rpa1,ISDFRPA,THC-RPA1} the application in molecular systems is still limited due to the $O(N^4)$ scaling of the generation for kernel of THC. 
While cubic scaling has been demonstrated by exploiting locality in basis functions or shell pairs,\cite{rpa_n3_1, rpa_n3_2,n3_gw,on3_gw_2} the formal scaling with respect to the number of basis functions remains quartic in those implementations.

To overcome the bottleneck of kernel generation, recently, we introduced the block tensor decomposition (BTD) algorithm, a dual-grid THC scheme that achieves formal $O(N^3)$ scaling in kernel construction.\cite{btd} By combining Hilbert space-filling curves with pivoted Cholesky decomposition, BTD generates compact and non-redundant interpolative grids, enabling efficient sparse mapping in real space. This algorithm has been implemented in the calculations of the Hartree-Fock (HF) exchange and the scaled opposite-spin MP2 (SOS-MP2) correlation.\cite{btd} In this work, we would like to extend the BTD framework to direct random phase approximation (dRPA) and $GW$ for molecule systems.

The motivation of this work is to presents the BTD-$GW$ algorithm, which integrates imaginary-time $GW$ formalism with BTD low-rank compression and Laplace transformation. This approach achieves formal $O(N^3)$ scaling, which is further reduced to near-quadratic scaling in practice by leveraging sparsity in the polarizability evaluated on interpolative grids. 

The article is organized as follows. In sections \ref{2.a} and \ref{2.b}, we briefly review the imaginary time $GW$ with the Laplace transform and the BTD algorithm, respectively. The section \ref{2.c} introduces the algorithm of $GW$ based on BTD. In this work, we implement both the one-shot $G_0W_0$ scheme and the eigenvalue self-consistent ev$GW$ scheme\cite{evGW} within the BTD framework. To assess the precision of the $GW$ method, the HOMO energies of the $GW$100 benchmark set are presented in Section~\ref{result}.A. The accuracy of the BTD-based algorithm is examined using the S66$\times$8 test set for non-covalent interactions.\cite{S66X8} Finally, the computational efficiency of BTD-RPA and BTD-$GW$ is demonstrated in Section~\ref{result}.B.

\section{Methodology}
The BTD-$GW$ algorithm achieves formal $O(N^3)$ scaling through a systematic strategy: 
(1) continuous real‐space quantities (e.g., $G(\bm{r},\bm{r}',\tau)$) are discretized on a compact set of interpolative grids, 
(2) these discrete representations are compressed via the block tensor decomposition (BTD), and 
(3) all subsequent many‐body operations (polarization, screening, self‐energy) are performed directly on the compressed tensors, avoiding explicit manipulation of high‐dimensional intermediates. 
For clarity, all notations are summarized in Table \ref{tab:small_mols}.

\begin{table}
\centering
\caption{Key notations used in the BTD-$GW$ formalism}
\begin{tabular}{c| c}
\hline
Notation&Description\\
\hline
$\mu,\ \nu,\ \lambda,\ \sigma$&indices for atomic orbitals (AOs)\\
$i,\ j,\ k,\ l$&indices of occupied orbitals\\
$a,\ b,\ c,\ d$&indices of virtual orbitals\\
$s,\ ,t\ ,u\ ,\ v$&indices of arbitrary ortbials\\
$g,\ h$&indices of dense grids\\
$K,\ L$&indices of interpolative grids\\
$M,\ N$&indices of auxiliary functions for RI\\
$\tau$&imaginary time\\
$i\omega$&imaginary frequency\\
$\bm{G}$&Green's function\\
$\bm{\Sigma}$&self-energy\\
$\bm{\Pi}$&polarizability\\
$\bm{W}$&dynamical coulomb screening\\
$\phi_u(\bm{r})$&value of orbital on grids\\
$\chi_M(\bm{r})$&value of auxiliary function on grids\\
$\epsilon^{QP}$&quasi-particle energy\\
$\bm{D}$&density matrix\\
$\bm{C}$&coefficient matrix of orbitals\\
\hline
\end{tabular}
\label{tab:small_mols}
\end{table}
\subsection{Imaginary-Time $GW$ Formalism with Laplace Transformation}\label{2.a}
Upon approximating the vertex functional, Hedin's equations simplify to a set of four coupled equations. To avoid divergences along the real-frequency axis, the calculations are performed on imaginary axis as follows.
\begin{subequations}
    \begin{align}
        &\bm{G}(i\omega) = [\bm{I} - \mathbf{h}_0 - \bm{\Sigma}(i\omega)]^{-1}\label{dyson_1}\\
        &\bm{P}(\tau) = -\bm{G}(\tau)\bm{G}(-\tau)\label{ndyson_1}\\
        &\bm{W}(i\omega) = \bm{v}[\bm{I} - \bm{v}\bm{P}(i\omega)]^{-1}\label{dyson_2}\\
        &\bm{\Sigma}(\tau) = -\bm{G}(\tau)\bm{W}(\tau)\label{ndyson_2}
    \end{align}
\end{subequations}
The one-electron Hamiltonian matrix $\mathbf{h}_0$ in eq. (2a) is the representation of the following operator $\hat{h}_0$ in the atomic orbital basis shown below: 
\begin{equation}
    \hat{h}_0 = -\frac{1}{2}\nabla^2-\sum_{a}\frac{Z_a}{\bm{R}_a}+\int\frac{\rho(\bm{r}')}{|\bm{r}-\bm{r}'|}\text{d}\bm{r}'
\end{equation}
The Hedin equations are solved in both the time and frequency domains, with the aim of avoiding complex convolution calculations. The Dyson equations of eq.(\ref{dyson_1}) and eq.(\ref{dyson_2}) are solved in the frequency domain. The polarization propagator and self-energy are built in the time domain as eq.(\ref{ndyson_1}) and eq.(\ref{ndyson_2}), respectively.


In the calculations of $GW$, the most time-consuming step is the calculation of dynamical coulomb screening $\bm{W}$. Using RI, polarizability $\bm{\Pi}$ and dynamical coulomb screening $\bm{W}$ are computed as follows.
\begin{subequations}
    \begin{align}
        &\Pi_{MN}(i\omega) = \sum_{ia}(M|ia)P_{ia}(i\omega)(ia|N)\\
        &W_{ia,ib}(i\omega) = \sum_{MN}(ia|M)[\bm{I}-\bm{\Pi}(i\omega)]^{-1}_{MN}(N|ib)
    \end{align}
\end{subequations}
Here $W_{ia,jb}(i\omega)$ is the dynamically screened Coulomb interaction in the orbital representation; while $P_{ia}(i\omega)$ is the polarization propagator in the orbital representation, which is defined by eq (5a)

In the zero-temperature limit, the irreducible polarization propagator $\bm{P}$ and the Green’s functions in orbital representation are calculated as below:
\begin{subequations}
\begin{align}
    &P_{ia}(\tau) = -G_a(\tau)G_i(-\tau)\\
    &G_a(\tau) = \exp(-\tau\epsilon_a) &&\qquad(\tau>0\text{ else }0)\\
    &G_i(-\tau) = -\exp(\tau\epsilon_i) &&\qquad(\tau>0\text{ else }0)
\end{align}
\end{subequations}

The dynamically screened Coulomb interaction $\bm{W}$, defined in eq.~(2c), exhibits different tensor ranks in different representations. In the molecular orbital basis, it is a fourth-order tensor $W_{ia,jb}(i\omega)$ . In real space representation, however, it reduces to a second-order tensor $W(\bm{r},\bm{r}',i\omega)$. This is the essential foundation for the subsequent application of low-rank decomposition as THC.

To perform transformations from imaginary time to imaginary frequency and sample points on imaginary axis, many algorithms have been developed for the zero temperature and finite temperature cases.\cite{irbasis, irsample, laplace1,cheb_sample, legendre_sample, sparse_sample} In this work, we employ a Laplace-transform based algorithm to connect imaginary-time and imaginary-frequency domains.\cite{laplace1, ls-mp22, GreenX} Specifically, cosine transforms are applied for the conversions $\bm{P}(\tau)\rightarrow\bm{P}(i\omega)$ and $\bm{W}(i\omega)\rightarrow\bm{W}(\tau)$.
\begin{subequations}
    \begin{align}
        &\bm{P}(\omega) = 2\int_0^\infty\cos(\omega\tau)\bm{P}(\tau)\text{d}\tau\\
        &\bm{W}(\tau) = \frac{1}{\pi}\int_0^\infty\cos(\omega\tau)\bm{W}(i\omega)\text{d}\omega
    \end{align}\label{cosine_trans}
\end{subequations}

The self-energy is divided into the odd part $\bm{\Sigma}^\text{s}$ and the even part $\bm{\Sigma}^\text{c}$. The sine transformation and cosine transformation are done for odd and even parts, respectively.
\begin{subequations}
    \begin{align}
    &\begin{aligned}
        \bm{\Sigma}(i\omega) = &2\int_0^\infty\cos(\omega\tau)\bm{\Sigma}^\text{c}(\tau)\text{d}\tau +\\ &2i\int_0^\infty\sin(\omega\tau)\bm{\Sigma}^\text{s}(\tau)\text{d}\tau
    \end{aligned}\\
    &\bm{\Sigma}^\text{c}(\tau) = -\frac{1}{2}[\bm{G}(\tau)+\bm{G}(-\tau)]\bm{W}(\tau)\\
    &\bm{\Sigma}^\text{s}(\tau) = -\frac{1}{2}[\bm{G}(\tau)-\bm{G}(-\tau)]\bm{W}(\tau)
    \end{align}\label{CS_trans}
\end{subequations}
These transformations are efficiently performed by introducing basis  
functions $\{\phi_\omega(x)\}$ and $\{\phi_\tau(x)\}$ to represent functions in the frequency and time domains, respectively: 
\begin{subequations}
\begin{align}
    &\phi_\omega(x) = \frac{2x}{x^2+\omega^2}\\
    &\phi_\tau(x) = \exp(-|\tau|x)
\end{align}
\end{subequations}
Transformation coefficients are calculated by least-squares minimization. 

The calculations of quasiparticle energies require the real-frequency self-energy, which is typically obtained via analytic continuation (AC) or contour deformation.\cite{contour_deform1, on3_gw_6} In $G_0W_0$, the quasi-particle energy of $u_\text{th}$ is calculated with Kohn-Sham orbital $\phi_u^{KS}$.
\begin{equation}
    \epsilon_u^{QP} = \text{Re}\langle\phi^{KS}_u|\hat{h}_0 + \hat{\Sigma}(\epsilon_u^{QP})|\phi^{KS}_u\rangle \label{QPEQ}
\end{equation}
In the eigenvalue self-consistent scheme, $\{\epsilon^{QP}_u\}$ are used to construct Green's function, and the Hedin's equations are solved iteratively.


The scaling of RI-$GW$ is $O(N^4)$. The computational bottleneck is the integral transformation of $(ia|M)$ and the calculation of $W_{ia,ib}$. Explicit construction of the full $W_{ia,jb}$ matrix would lead to $O(N^5)$ scaling.

To avoid this quartic and quintic scaling, instead of the orbital representation,  the construction of $W_{ia,jb}$ is reformulated in real space, which provides a more natural framework for exploiting spatial sparsity and enabling low-rank approximations. This strategy avoids the explicit construction of the $W_{ia,jb}$ matrix and the costly orbital integral transformations. Therefore, 
the Green's function and the polarization propagator in real space are represented as follows.
\begin{subequations}
    \begin{align}
        &G(\bm{r},\bm{r'},\tau) =\left\{ 
        \begin{array}{cc}
        &\sum\limits_{i\in \text{occ}}\phi_i(\bm{r})\phi_i^*(\bm{r}')\exp(-\tau\epsilon_i);\ \tau<0\\
        &-\sum\limits_{a\in\text{vir}}\phi_a(\bm{r})\phi_a(\bm{r}')\exp(-\tau\epsilon_a);\ \tau>0
        \end{array}\right.\\
        &P(\bm{r},\bm{r}',\tau) = -G(\bm{r},\bm{r'},\tau)G(\bm{r},\bm{r'},-\tau)
    \end{align}\label{rs-GW}
\end{subequations}

It is noted that in this work, $\bm{P}$ generically denotes the polarization propagator (density–density response function). In real space representation, we compute $P(\bm{r},\bm{r}',\tau)$ via eq.~(10b); after compression onto interpolative grids it becomes $P_{KL}(\tau)$. In the molecular orbital representation introduced in eq.~(5a), it is denoted as $P_{ia}$, and its frequency-domain form $P_{ia}(i\omega)$ is obtained by Fourier transform of $P_{ia}(\tau)$. Projection onto the RI auxiliary basis gives the polarizability matrix $\Pi_{MN}(i\omega)$, which appears in both the screened interaction and the RPA correlation energy discussed in the section \ref{2.c}. 
Thereafter, the BTD algorithm introduced in the section \ref{2.b} is used in the calculations of polarizability.
\subsection{Block tensor decomposition}\label{2.b}
In the BTD, electronic integrals are approximated by the interactions of discrete point charges at $\{\bm{r}_g\}$.\cite{btd} These point charges can be contracted to a set of interpolative charges $\{\bm{r}_K\}$ with fewer grids by using overlap fitting.
\begin{subequations}
    \begin{align}
        &(\mu\nu|\lambda\sigma) = \sum_{KL}X_{\mu K}X_{\nu K}V_{KL}X_{\lambda L}X_{\sigma L}\\
        &V_{KL} = \sum_{gh}\xi_{Kg}\xi_{Lh}V_{gh}\\
        &\xi_{Kg}=\sum_LS^{-1}_{KL}\sum_{\mu\nu}Q_{g}^{\mu\nu}Q_{L}^{\mu\nu}\\
        &S_{KL}=\sum_{\mu\nu}Q^{\mu\nu}_KQ^{\mu\nu}_L\label{SKL}\\
        &Q^{\mu\nu}_g = X_{\mu g}X_{\nu g};\ Q^{\mu\nu}_K = X_{\mu K}X_{\nu K}
    \end{align}
\end{subequations}
A set of continuous charges is introduced by RI. The point-point interaction $V_{KL}$ is decomposed into a contraction of two point–continuum interaction matrices via eq.~(\ref{CD-int}), thereby circumventing the Coulomb cusp $\lim_{r\rightarrow 0}1/r$.
\begin{subequations}
    \begin{align}
        &V_{KL} = \sum_{M}B_{KM}B_{LM}\\
        &B_{KM} = \sum_{g}\xi_{Kg}B_{gM}\label{time_max}\\
        &B_{gM} = \sum_{N}(M|N)^{-1/2}\int\frac{\chi_M(\bm{r})}{|\bm{r} - \bm{r}_g|}\text{d}\bm{r}
    \end{align}\label{CD-int}
\end{subequations}
BTD employs a dual-grid scheme to replace the computation of 3-center 2-electron (3c2e) integrals with 2-center 1-electron (2c1e) integrals. Compared to other THC schemes for molecular systems, such as least-squares THC (LS-THC),\cite{thc1} THC-RI,\cite{isdf3} and separable density fitting,\cite{isdf-rpa1} BTD exhibits formal cubic scaling with both system size and number of basis functions.

To generate the interpolative points in an efficient and compact way, a scheme based on Hilbert sorting/Octree and pivoted Cholesky decomposition was proposed. This scheme can generate non-redundant grids in a nearly linear-scaling way. The number and distribution of grids are closely related to accuracy and efficiency. To balance efficiency and accuracy, the parameters for dense grids and interpolative grids are optimized by a parameter adaptive differential evolution algorithm named JADE, with details provided in our previous work.\cite{btd}

The most time-consuming step of BTD is eq. (\ref{time_max}) with a cost of $O(N_\text{aux}N_KN_g)$. To reduce time consumption, the sparsity of real space is explored. The sparse map\cite{smap1, smap2, smap3} between massive grids $\{\bm{r}_g\}$ and interpolative grids $\{\bm{r}_K\}$ is determined by 
\begin{equation*}
    L(\bm{r}_g\rightarrow\bm{r}_K) = L(\bm{r}_g\rightarrow\chi_\mu)\cup L(\chi_\mu\rightarrow\bm{r}_K)
\end{equation*}
The cost is reduced to $O(N_\text{aux}N_K)$ as the size of the system grows. The bottleneck is solving the linear equation for overlap fitting with a time scale $O(N_K^3)$. In the tests of 1D and 3D systems, BTD shows a scaling between $O(N^2)$ and $O(N^3)$. The implementations of BTD in HF exchange and SOS-MP2 correlation have shown excellent performance. 
\subsection{Formal $O(N^3)$ scaling  BTD-$GW$ Algorithm}\label{2.c}
In this section, by applying BTD, all intermediate quantities are directly computed and compressed on the set of interpolative grids $\{\bm{r}_K\}$. Specifically, the continuous Green's function $G(\bm{r},\bm{r}',\tau)$ of eq.~(10a) is sampled directly on the interpolative grid points to yield the compressed representation $G_{KL}(\tau)$. From this, the polarization propagator $P_{KL}(\tau)$ is constructed via $P_{KL}(\tau) = G_{KL}(\tau)G_{KL}(-\tau)$, mirroring the real space relation in eq.~(10b). The dynamically screened Coulomb interaction $W$ is similarly compressed as $W_{KL}$. This allows us to efficiently compute the self-energy in the atomic-orbital representation without explicitly constructing the fourth-order tensor $W_{ia,jb}$.

In our implementation, the self-energy is divided into the static and dynamic parts, a common practice in $GW$ calculations. The static part, $\bm{\Sigma}_x(0)$ is the frequency-independent Hartree-Fock exchange potential at zero frequency, which can be calculated by chain of spheres for exchange (COSX).\cite{Neese_2009, Izsak2011, Izsak2013, Helmich-Paris2021, aCOSX} The dynamic part $\bm{\Sigma}_c(i\omega)$, also known as the correlation self-energy, encapsulates all the frequency-dependent correlation effects arising from the screened interaction $W$. To obtain the dynamic correlation self-energy $\Sigma_c(i\omega)$, the Green's function is first evaluated on the interpolative grids as $G_{KL}(\tau)$ by direct sampling of eq.~\eqref{rs-GW}. The dynamically screened Coulomb interaction $W_{KL}(i\omega)$ is then constructed on the same grids as follows:
\begin{subequations}
    \begin{align}
        &W_{KL}(i\omega) = \sum_{MN}B_{KM}B_{LN}[\bm{I} - \bm{\Pi}(i\omega)]^{-1}_{MN}\\
        &\Pi_{MN}(i\omega) = \sum_{KL}B_{KM}B_{LN}P_{KL}(i\omega)
    \end{align}\label{RS-WPI}
\end{subequations}
\indent The $\bm{\Sigma}_c(i\omega)$ can be represented in the orbital representation $\Sigma_{u,c}(\tau)$ and in the basis representation $\Sigma_{\mu\nu,c}(\tau)$ via the equations below.
\begin{subequations}
\begin{align}
    &\Sigma_{u,c}(\tau) = \sum_{\mu\nu}C_{u\mu}C_{u\nu}\Sigma_{\mu\nu,c}(\tau)\\
            &\Sigma_{\mu\nu,c}(\tau) = -\sum_{KL}X_{\mu K}X_{\nu L}G_{KL}(\tau)[W_{KL}(\tau) - V_{KL}]
\end{align}\label{SE-build}
\end{subequations}

\noindent Here, the term $V_{KL}$ is subtracted to avoid double-counting the static exchange already included in the Hartree-Fock exchange.

In this work, the analytic continuation of $\bm{\Sigma}_c(i\omega)$ to the real-frequency axis is performed with the Pade approximation. The Pade coefficients are obtained by fitting $\bm{\Sigma}_c(i\omega)$ on the imaginary-frequency grid points, and the resulting rational function is evaluated on the real axis to yield $\bm{\Sigma}_c(\omega)$. The nonlinear equation eq.(\ref{QPEQ}) of quasi-particle energy is solved by linear search. With the dynamically screened Coulomb interaction $W_{KL}$ available on the interpolative grids, the quasiparticle energies $\epsilon_u^\mathrm{QP}$ are determined, completing one cycle of the $GW$ iteration.

Separately, once the polarizability $\Pi_{MN}(i\omega)$ is obtained via eq.~(13b), the direct random phase approximation (dRPA) correlation energy can be evaluated using the adiabatic-connection fluctuation-dissipation theorem (ACFDT).
\begin{equation}
    E^\text{RPA}=\frac{1}{2\pi}\int_{0}^{\infty}\ln(\text{det}[\bm{I}-\bm{\Pi}(i\omega)])+\text{Tr}[\bm{\Pi}(i\omega)]\text{d}\omega\label{RPA_EF}
\end{equation}
Our BTD-dRPA implementation captures long-range dispersion interactions with an $O(N^3)$ scaling, offering computational efficiency superior to many post-SCF methods. We implemented and validated the random phase approximation within the BTD framework (BTD-RPA), which naturally served as a precursor and testing ground for our BTD-$GW$ development. This established the cubic-scaling foundation subsequently extended to the full $GW$ formalism.

The scaling of BTD-RPA is formally cubic and the computational bottleneck is the building of $\Pi_{MN}(i\omega)$ in eq.(\ref{RS-WPI}b). To reduce this cost, a prescreening strategy is introduced that takes advantage of the sparsity of the polarization propagator $P_{KL}(\tau)$ in the time domain. Only grid pairs $(K,L)$ with non-negligible $P_{KL}(\tau)$ values are retained for the correlation energy calculation, as negligible pairs can be screened out. To enable this screening efficiently, the atomic orbital representation of the Green's functions, as defined in eqs. (16a) and (16b), is employed to naturally exploit their spatial locality.
\begin{subequations}
\begin{align}
&\begin{aligned}
 G^>_{KL}(\tau) &= -\sum_{\mu\nu}X_{\mu K}X_{\nu L}(\sum_{a\in\text{vir}}C_{a\mu}C_{a\nu}\exp(-\tau\epsilon_a))\\
 &= \sum_{\mu\nu}X_{\mu K}X_{\nu L} G^>_{\mu\nu}(\tau) \quad (\tau > 0)
\end{aligned}\\
&\begin{aligned}
 G^<_{KL}(-\tau) &= \sum_{\mu\nu}X_{\mu K}X_{\nu L}(\sum_{i\in\text{occ}}C_{i\mu}C_{i\nu}\exp(\tau\epsilon_i))\\
 &= \sum_{\mu\nu}X_{\mu K}X_{\nu L} G^<_{\mu\nu}(-\tau) \quad (\tau > 0)
\end{aligned}
\end{align}
\end{subequations}
Here, $C_{a\mu}$ is the molecular orbital coefficient. Because of the limitation relationship $\lim_{\tau\rightarrow0}G^<_{\mu\nu}(\tau) = D_{\mu\nu}$, $G_{KL}^{<}(-\tau)$ has a similar local property to the density matrix $D_{\mu\nu}$. The prescreening is performed by estimating $X_{\mu K} X_{\nu L}\max_{KL}(\bm{D})$. 
To enable efficient prescreening, a representative atomic orbital index $\mathrm{p}_K$ is precomputed for each interpolative grid block $K$. This index points to the orbital within the sparse map of block $K$ that has the largest row norm in the density matrix $\mathbf{D}$. The maximum possible density matrix element between a pair of blocks $(K, L)$ can then be efficiently estimated as $D_{\mathrm{p}_K, \mathrm{p}_L}$. This estimation procedure is outlined in the following pseudo-code:
\begin{equation}
\begin{aligned}
    &\textbf{REQUIRED: }L(\mu\rightarrow\bm{r_K}),\ L(\mu\rightarrow\bm{r_L}),\ \bm{D}\\
    &\textbf{ENSURE: }\max(D_{\mu\nu})\\
    &\text{set max\_val}=0;\ \text{p}_K = -1\\
    &\textbf{FOR }\mu\in L(\mu\rightarrow\bm{r}_K):\\
    &\quad \textbf{IF }\max(\bm{D}[\mu,:])>\text{max\_val}:\\
    &\qquad\text{max\_val} = \max(\bm{D}[\mu,:]);\ \text{p}_K = \mu\\
    &\textbf{ENDFOR}\\
    &\text{determine }\text{p}_L \text{ for }\bm{r}_L\\
    &\max{}_{KL}(\bm{D}) = D_{\text{p}_K,\text{p}_L}
\end{aligned}
\end{equation}
Here, the index $\text{p}_K$ can be determined before the computation of $\max_{KL}(\bm{D})$. Using the sparse maps between interpolative grids, the scaling of building $\Pi_{MN}$ is reduced to $O(N_\text{aux}^2N_K)$.
A key challenge in the $GW$ formalism arises from the long-range nature of the screened Coulomb interaction $\mathbf{W}(\tau)$, which prevents the direct application of the spatial screening schemes used in dRPA. Consequently, the computation of the self-energy $\Sigma_u(\tau)$ becomes the dominant cost, scaling as $O(N_K^2 N_\text{bas})$. 


As illustrated in Figure 1, the formal $O(N^3)$ scaling of the ev$GW$ algorithm arises from the collective scaling behavior of its constituent components. The calculation scheme exhibits an overall $O(N^3)$ scaling. This formal cubic scaling is a direct result of the BTD compression, which ensures that all other key steps, such as building $\Pi_{MN}$ and inverting the dielectric matrix, also scale no worse than $O(N^3)$.

The calculation procedure of BTD-RPA and BDT-$GW$ is summarized in Figure \ref{bse_procedure}. 
Starting from a mean-field (HF or DFT) reference, the workflow follows the systematic compression strategy outlined above: molecular orbitals and the density matrix are projected onto interpolative grids, continuous real space quantities (e.g., $G$, $P$) are discretized and compressed via BTD, and all subsequent many-body operations are performed directly on the compressed tensors. This integrated pipeline enables the formal $O(N^3)$ scaling demonstrated in the following section.

\begin{figure}[h]
\begin{equation*}
    \begin{aligned}
        &\textbf{REQUIRED: }X_{\mu K},\ B_{KM},\ \{\phi_u\},\ \{\epsilon^{KS}_u\}\\
        &\textbf{ENSURE: }W_{KL}(0),\ \{\epsilon_u^{QP}\}\\
        &\text{let }\epsilon_u^{QP} = \epsilon^{KS}_u\\
        &\text{calculate $\bm{F} = \bm{h}_0+\bm{\Sigma}_x$ with $\{\phi_u\}$ using RI-COSX}\\
        &\textbf{DO}\\
        &\quad\text{calculate Green's function using eq.(\ref{rs-GW}a)}\\
        &\quad\text{calculate $\bm{P}(\tau)$ on interpolative grids using eq.(\ref{rs-GW}b)}\\
        &\quad\text{perform the cosine transform for }P_{KL}\text{ using eq.(\ref{cosine_trans}a)}\\
        &\quad\textbf{FOR }\text{block pairs with no zero $P_{KL}$}\\
        &\qquad\text{calculate polarizability }\Pi_{MN}(i\omega)\text{ using eq.(\ref{RS-WPI}b)}\\
        &\quad\textbf{END FOR}\\
        &\quad\text{calculate }[\bm{I}-\bm{W}(i\omega)]^{-1}\\
        &\quad\text{perform the cosine transform for }[\bm{I}-\bm{W}(i\omega)]^{-1}\text{ as eq.(\ref{cosine_trans}b)}\\
        &\quad\text{calculate self-energy as }\Sigma^\text{c}_{\mu\nu}(\tau)\text{ and }\Sigma^\text{s}_{\mu\nu}(\tau)\\
        &\quad\text{perform sine and cosine transforms for $\bm{\Sigma}$ using eq.(\ref{CS_trans})}\\
        &\quad\text{calculate Pade approximation of }\Sigma_i(\omega)\\
        &\quad\text{calculate $\epsilon^{QP}_u$ using eq.(\ref{QPEQ})}
        \\&\textbf{WHILE }\text{don't converge}\\
        &\text{calculate dynamical coulomb screening with }\omega=0
    \end{aligned}
\end{equation*}
    \caption{Workflow of the BTD-ev$GW$ algorithm}
    \label{evGW_al}
\end{figure}

\begin{figure}
    \centering
    \includegraphics[width=1.0\linewidth]{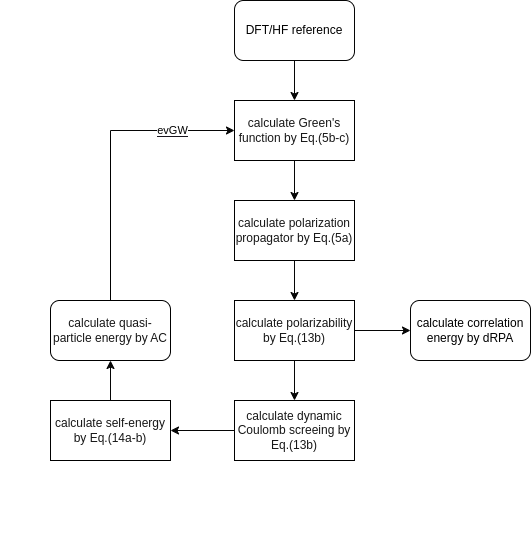}
    \caption{The procedure of BTD-RPA and BTD-$GW$}
    \label{bse_procedure}
\end{figure}

\section{Results and discussions}\label{result}
\subsection{Accuracy Assessment: $GW$100 and S66 Benchmarks}
 The algorithm is implemented in the developing version of XEDA software.\cite{tang_xeda_2021} The calculations of electron integrals are supported by Libxint which is the electron integral library of XEDA. All calculations were performed on Intel(R) Xeon(R) Gold 6226R CPU @ 2.90GHz. The cc-pVDZ basis set was employed, and DFT calculations were supported by Libxc.\cite{Lehtola2018} The cc-pVDZ-RI basis was used for auxiliary functions.

The accuracy of the BTD-based $GW$ method is assessed using the $GW$100 benchmark set,\cite{gw100, gw100_bg0w0, gw100_paw, gw100_west, gw100_sto} which comprises 55 molecules containing the elements C, H, O, N, S, P, F, and Cl. The quasiparticle energies are computed using $GW$/cc-pVTZ with six different reference states: $\omega$B97X-D, CAM-B3LYP, Hartree–Fock, M06-2X, PBE0, and B3LYP. The calculations of the no-iterative $G_0W_0$ and the approximate iterative scheme ev$GW$ are performed. 
\begin{figure*}
    \centering
     \subfloat[$\omega$B97X-D]{
        \includegraphics[width=0.35\linewidth]{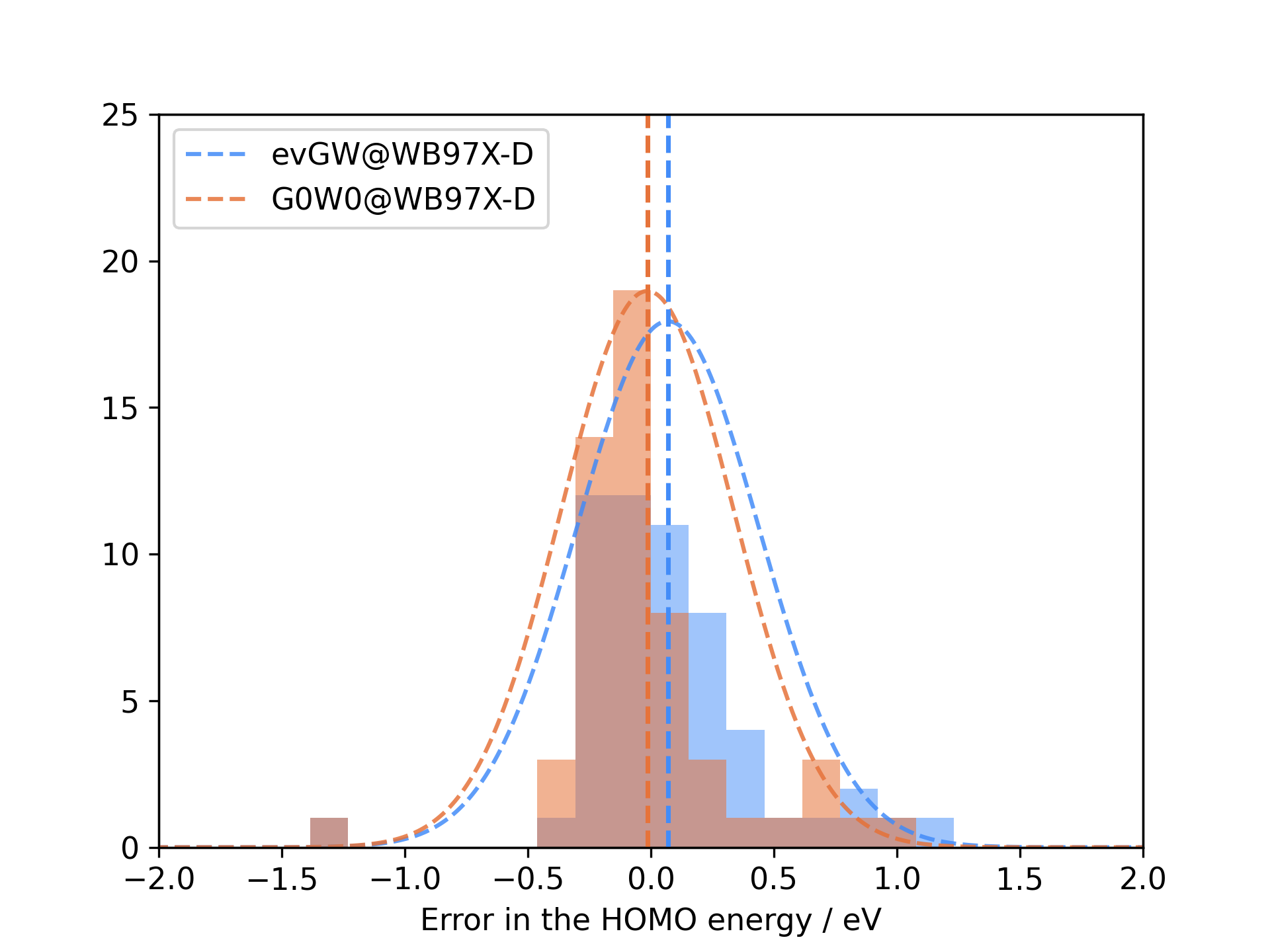}\label{GW100_A}
      }
      \subfloat[CAM-B3LYP]{
        \includegraphics[width=0.35\linewidth]{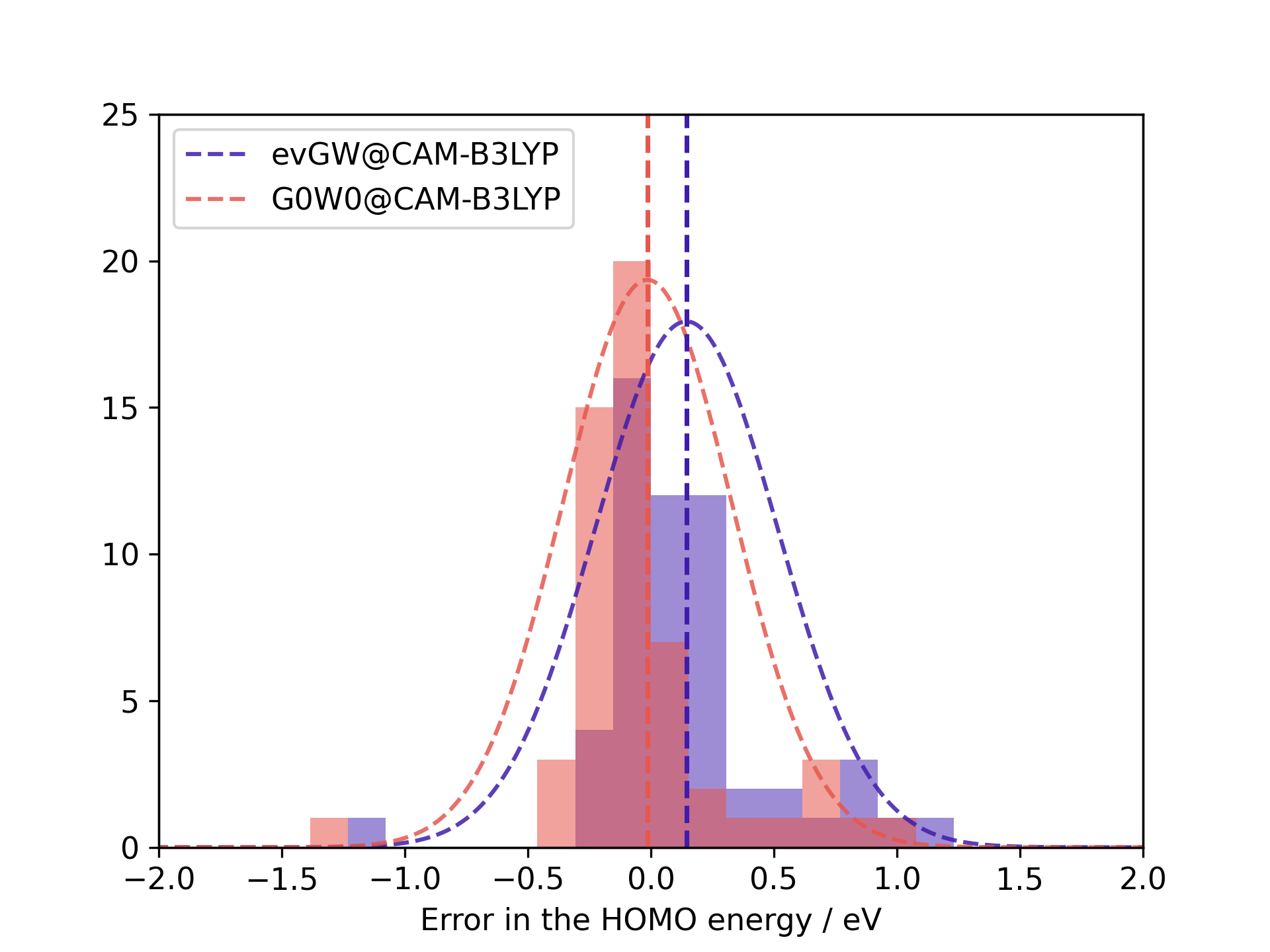}\label{GW100_B}
      }
      \subfloat[HF]{
        \includegraphics[width=0.35\linewidth]{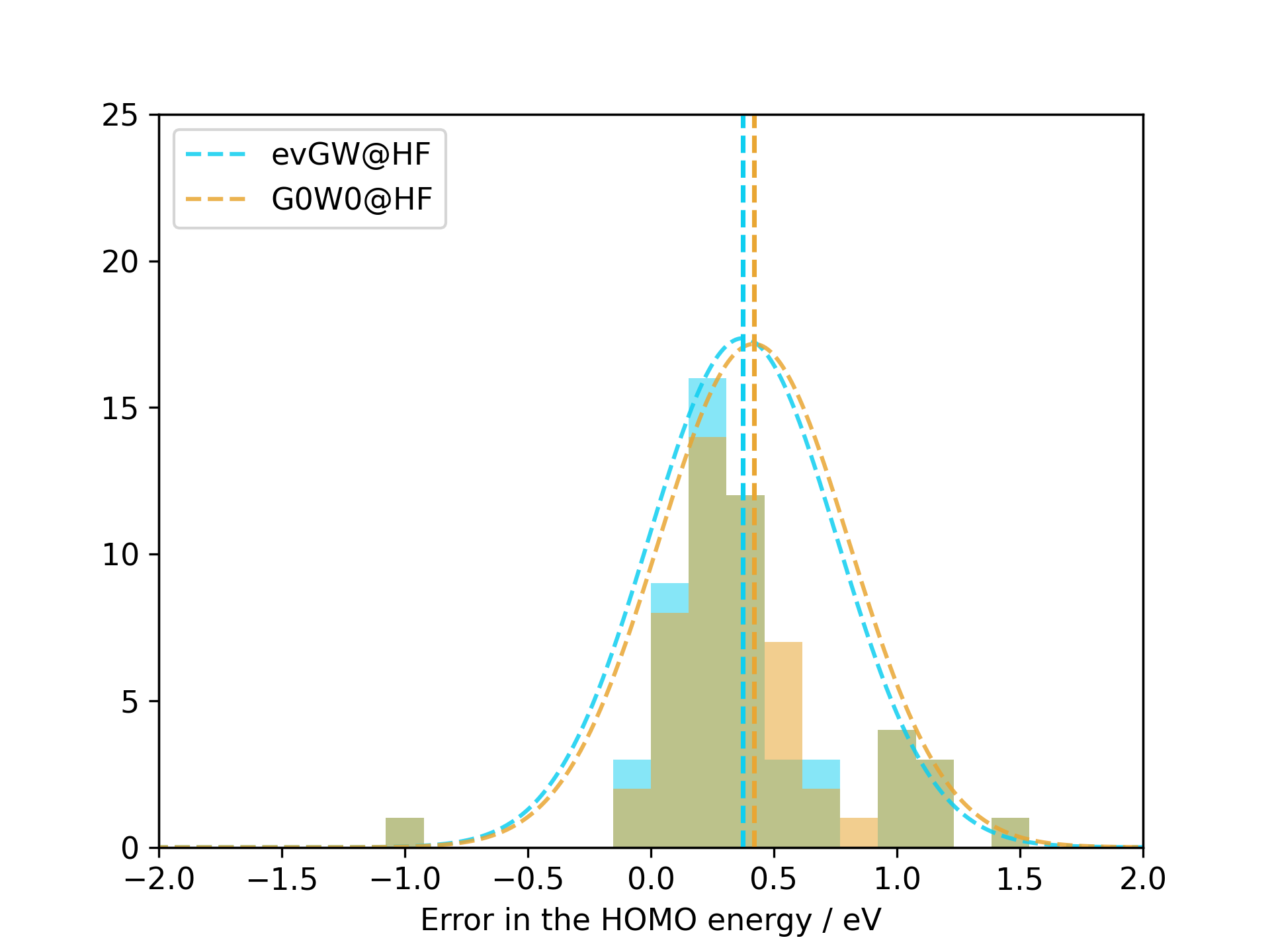}
      }\\
      \subfloat[M06-2X]{
        \includegraphics[width=0.35\linewidth]{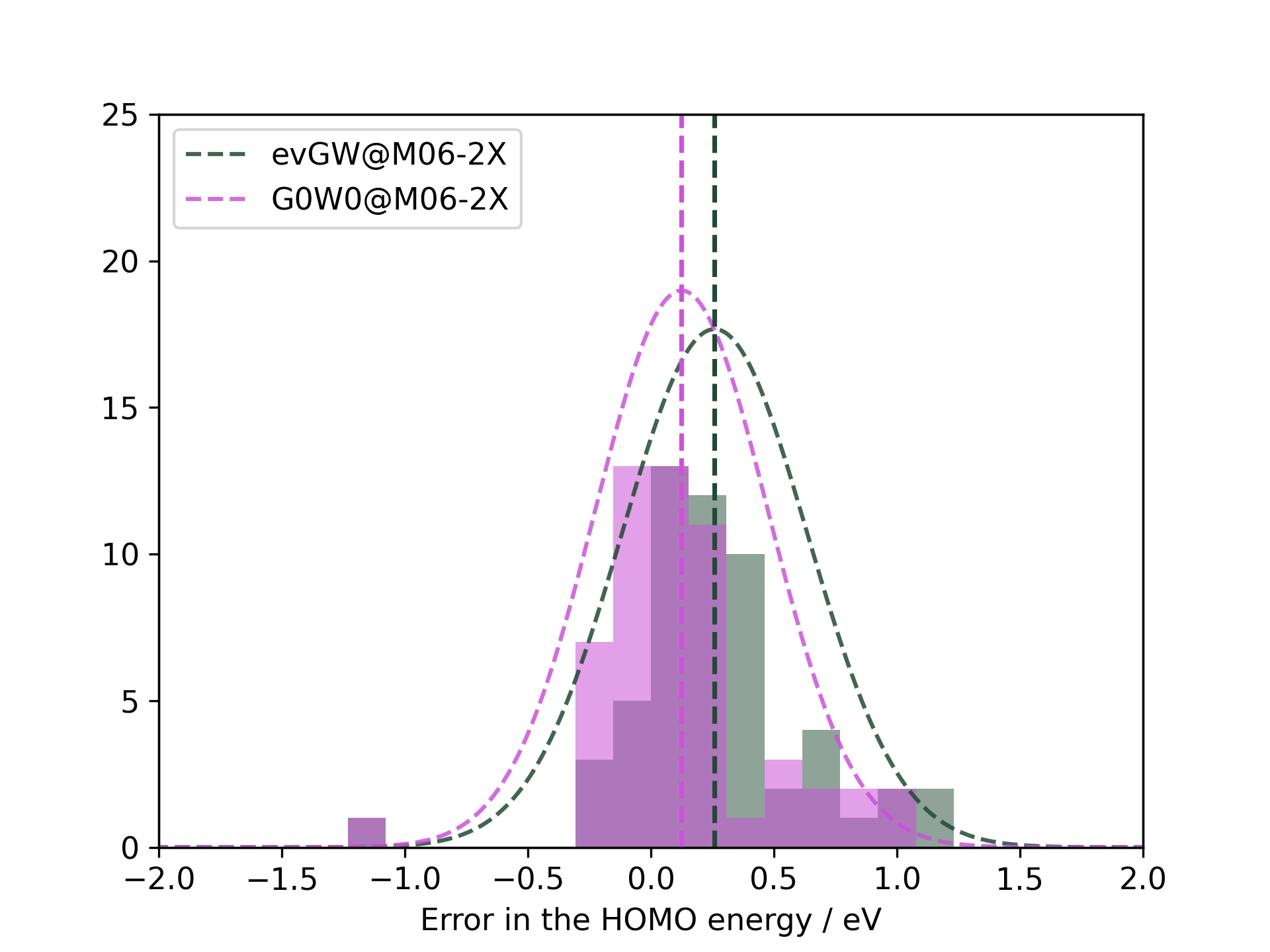}
      }
      \subfloat[PBE0]{
        \includegraphics[width=0.35\linewidth]{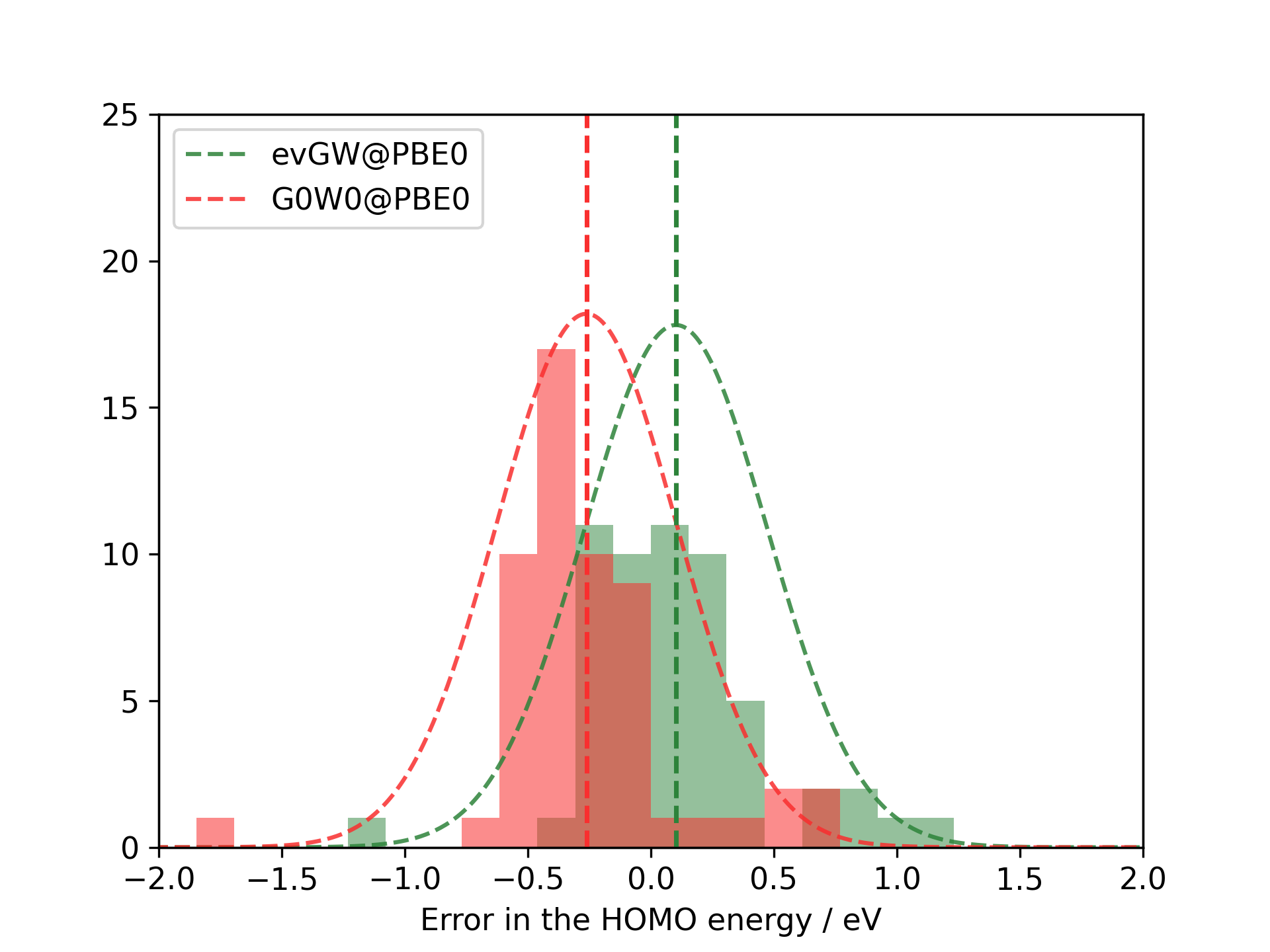}
      }
      \subfloat[B3LYP]{
        \includegraphics[width=0.35\linewidth]{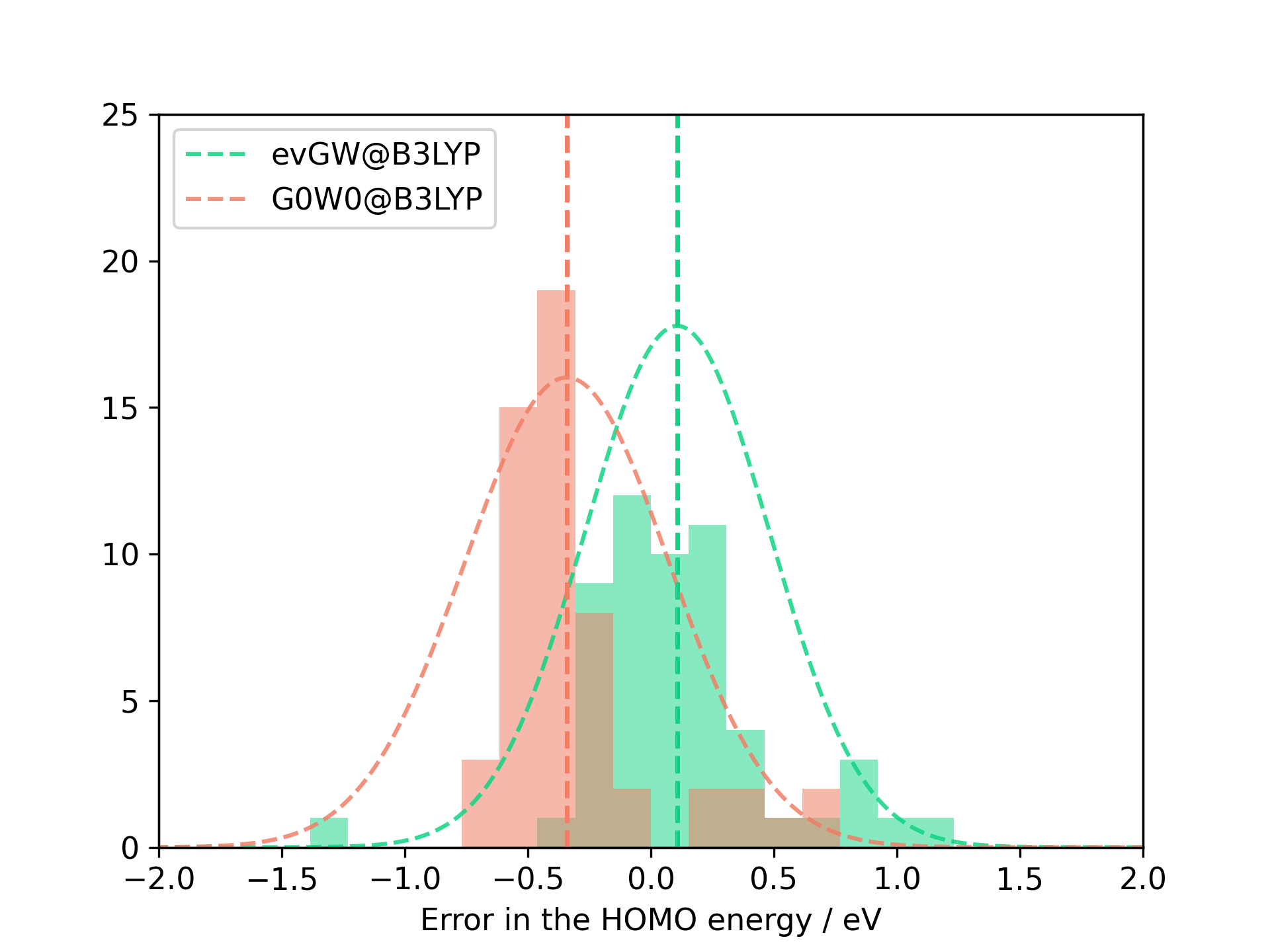}
      }
      \caption{Deviations of $G_0W_0$ and ev$GW$ HOMO energies from experimental ionization potentials for the $GW$100 set}
      \label{GW100}
\end{figure*}

Figure ~\ref{GW100} shows the comparison between the computed quasiparticle HOMO energies from $GW$ and the experimental vertical ionization potentials (IPs). The original data are shown in Table S1 and Table S2 in the supporting information. The results of $G_0W_0$ and ev$GW$ calculations exhibit a mean absolute error of approximately 1.0 eV. The largest error of about 1.5 eV is observed for tetracarbon (C$_4$). This molecule possesses a significant multi-reference character, with four unpaired electrons coupling to form a singlet ground state. Such a multi-reference character is fundamentally inaccessible to a single-determinant description, whether from restricted HF or KS-DFT. It is important to note that the ev$GW$ approach, while self-consistent in the eigenvalues, does not remedy this fundamental issue. As it starts from and iterates upon a single-determinant framework, it primarily captures dynamic correlation and cannot recover the strong static correlation present in C$_4$. Consequently, the initial error in the reference state propagates into an inaccurate polarization propagator $\bm{P}$, leading to the observed deviation.

A detailed analysis of Figure 3 reveals that the effect of the ev$GW$ correction varies systematically and significantly with the choice of starting reference. For the hybrid functionals B3LYP and PBE0, the ev$GW$ procedure consistently shifts the IPs upward relative to $G_0W_0$, leading to a marked improvement. This systematic improvement stems from the fact that the starting Kohn-Sham HOMO-LUMO gap is typically underestimated in these functionals due to their insufficient fraction of exact exchange. The ev$GW$ procedure self-consistently updates the quasiparticle energies in the Green's function $G$. Through this iteration, the fundamental gap is progressively opened toward the quasi-particle gap, thereby incorporating a more realistic exchange-correlation effect and shifting the IPs upward, which yields better agreement with the experiment data. This mechanism is conceptually analogous to that in the renormalized single (RS) Green function and the renormalized single-excitation (rSE) in the RPA.\cite{rsGW_1, rsGW_2, bse4mol_4, rseRPA_1, rseRPA_2, rseRPA_3}

In contrast, for range-separated functionals $\omega$B97X-D and CAM-B3LYP, the $G_0W_0$ starting point is already accurate, because they approximate the long-range exchange in the $GW$ self-energy, as shown in Figures \ref{GW100_A} and \ref{GW100_B}. In such cases, the additional ev$GW$ correction can over-adjust the already reasonable gaps, sometimes degrading the performance. The meta-hybrid M06-2X shows the similar behavior, where ev$GW$ correction sometimes degrades the result, suggesting a mismatch between its complex functional form and the specific correlation effects addressed by the ev$GW$ iteration.

Finally, for HF reference, which lacks dynamic correlation and overestimates the gap, ev$GW$ provides the limited improvement because it does not fundamentally rectify the missing correlation in the reference orbitals. 

\subsection{Performance Analysis: Scaling and Efficiency}
\begin{figure}[h]
    \centering
     \subfloat[Training process]{
        \includegraphics[width=1.0\linewidth]{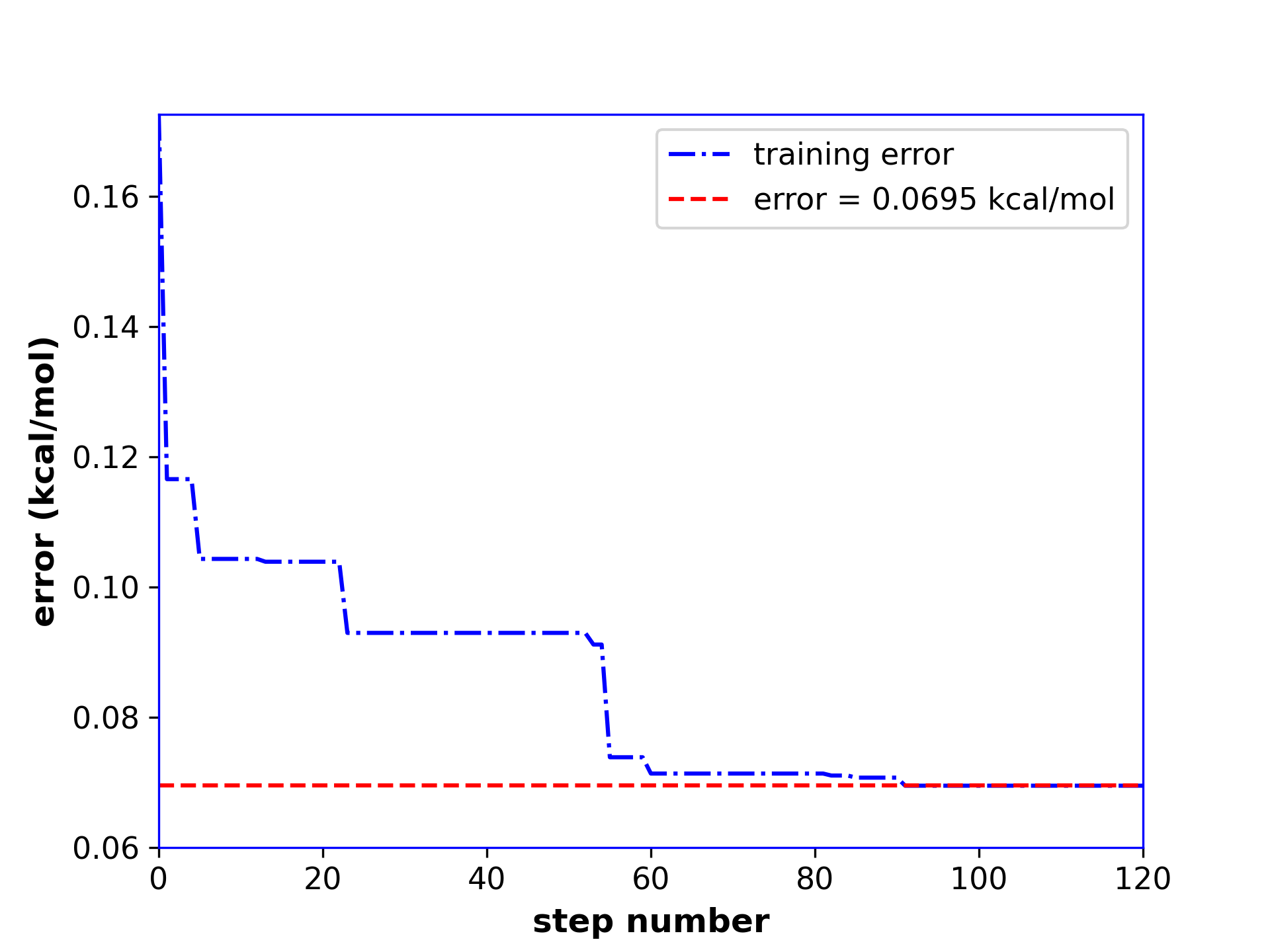}
      }\\
      \subfloat[Testing results on S66$\times$8]{
        \includegraphics[width=1.0\linewidth]{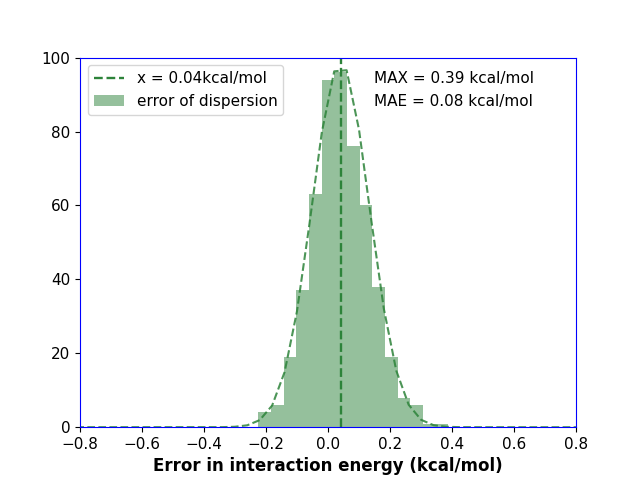}
      }
      \caption{(a) Training process on S66 test, (b) Error distribution of BTD-RPA compared to RI-RPA on S66$\times$8 with the BLYP reference}
      \label{s66x8}
\end{figure}
\begin{figure*}
    \centering
     \subfloat[Scaling of glycine chain's calculations]{
        \includegraphics[width=0.5\linewidth]{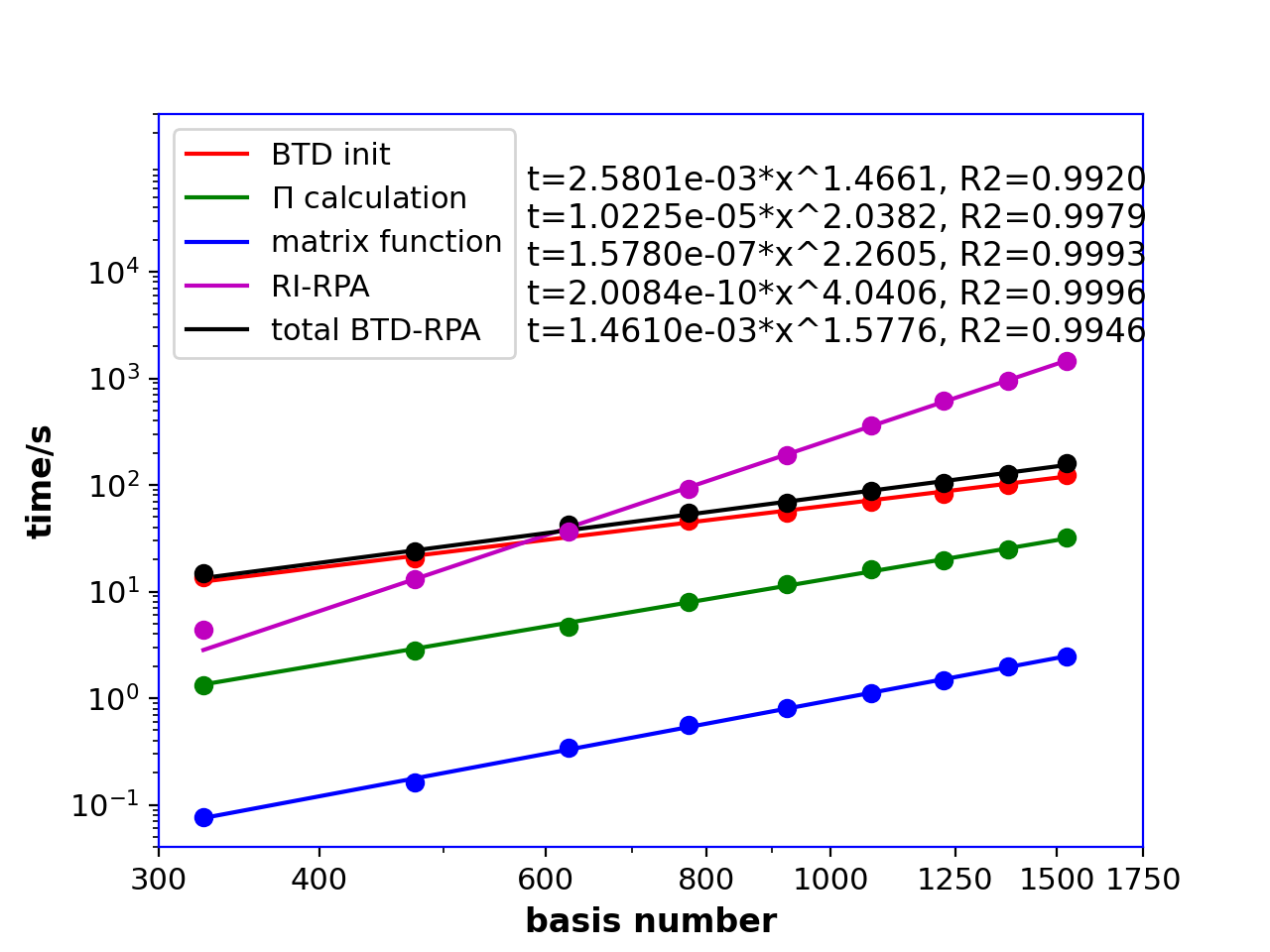}
      }
    \subfloat[Scaling of water cluster's calculations]{
        \includegraphics[width=0.5\linewidth]{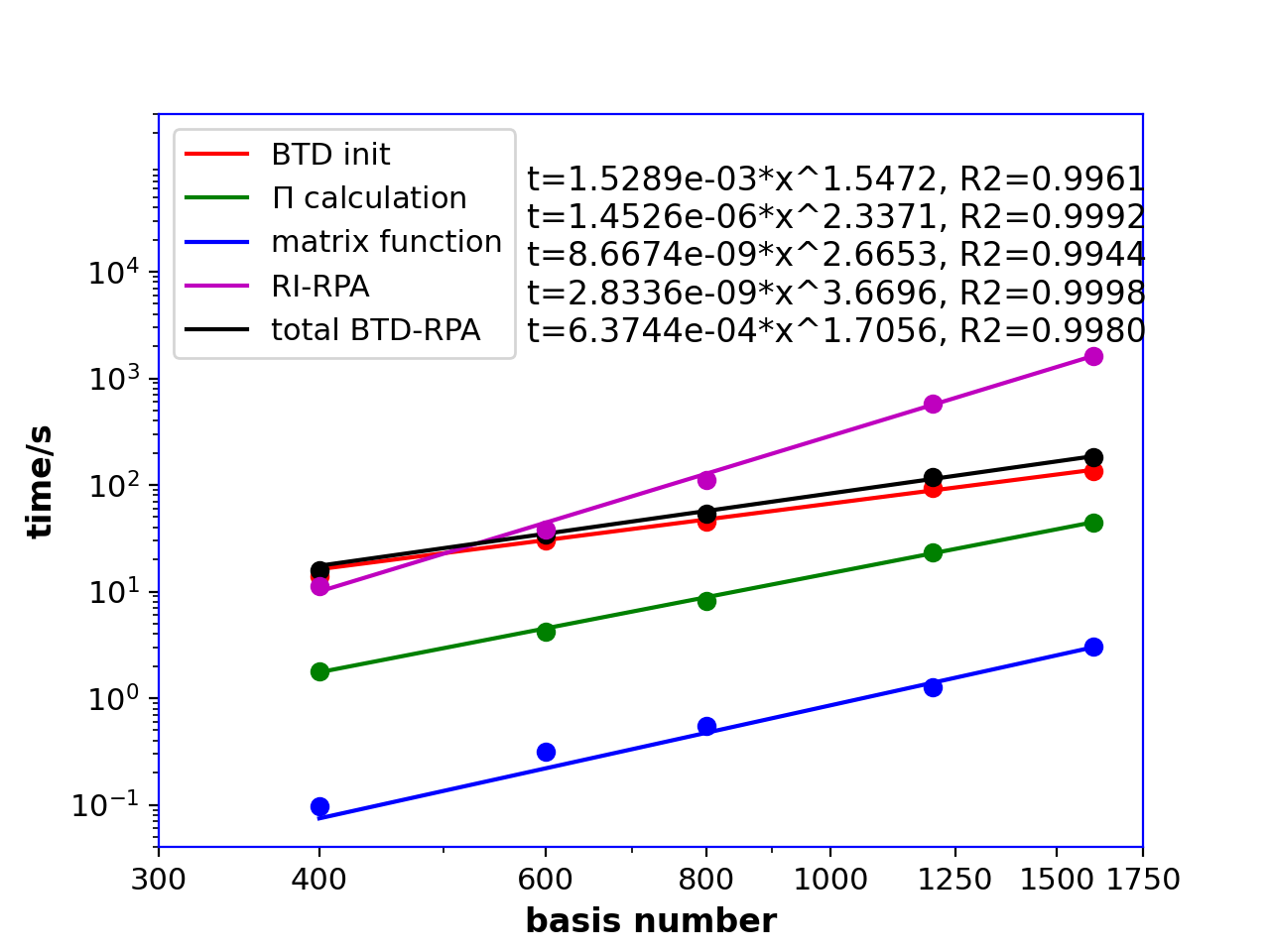}
      }\\
      \subfloat[Biase of glycine chain's calculations]{
        \includegraphics[width=0.5\linewidth]{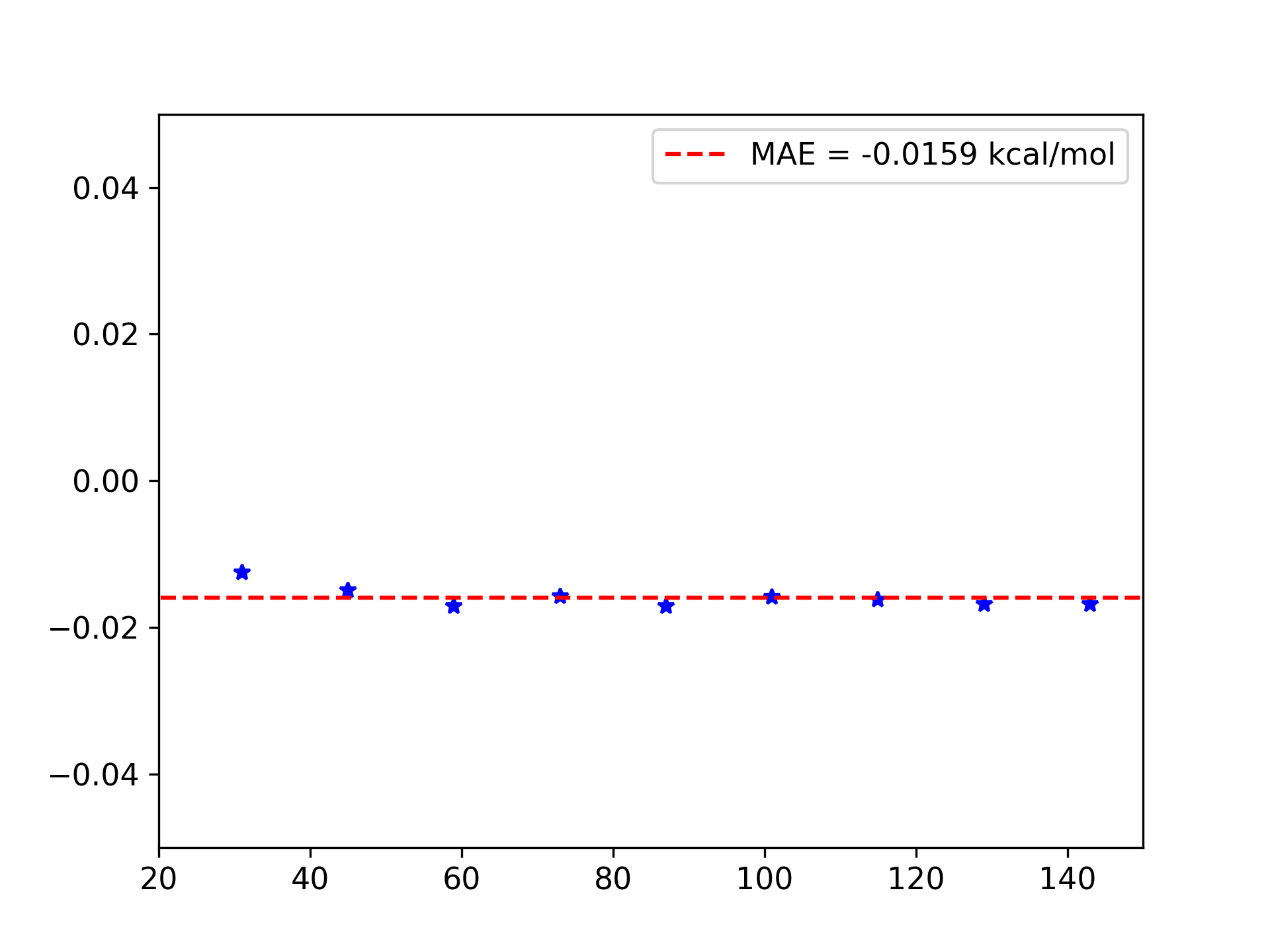}
      }
    \subfloat[Biase of water cluster's calculations]{
        \includegraphics[width=0.5\linewidth]{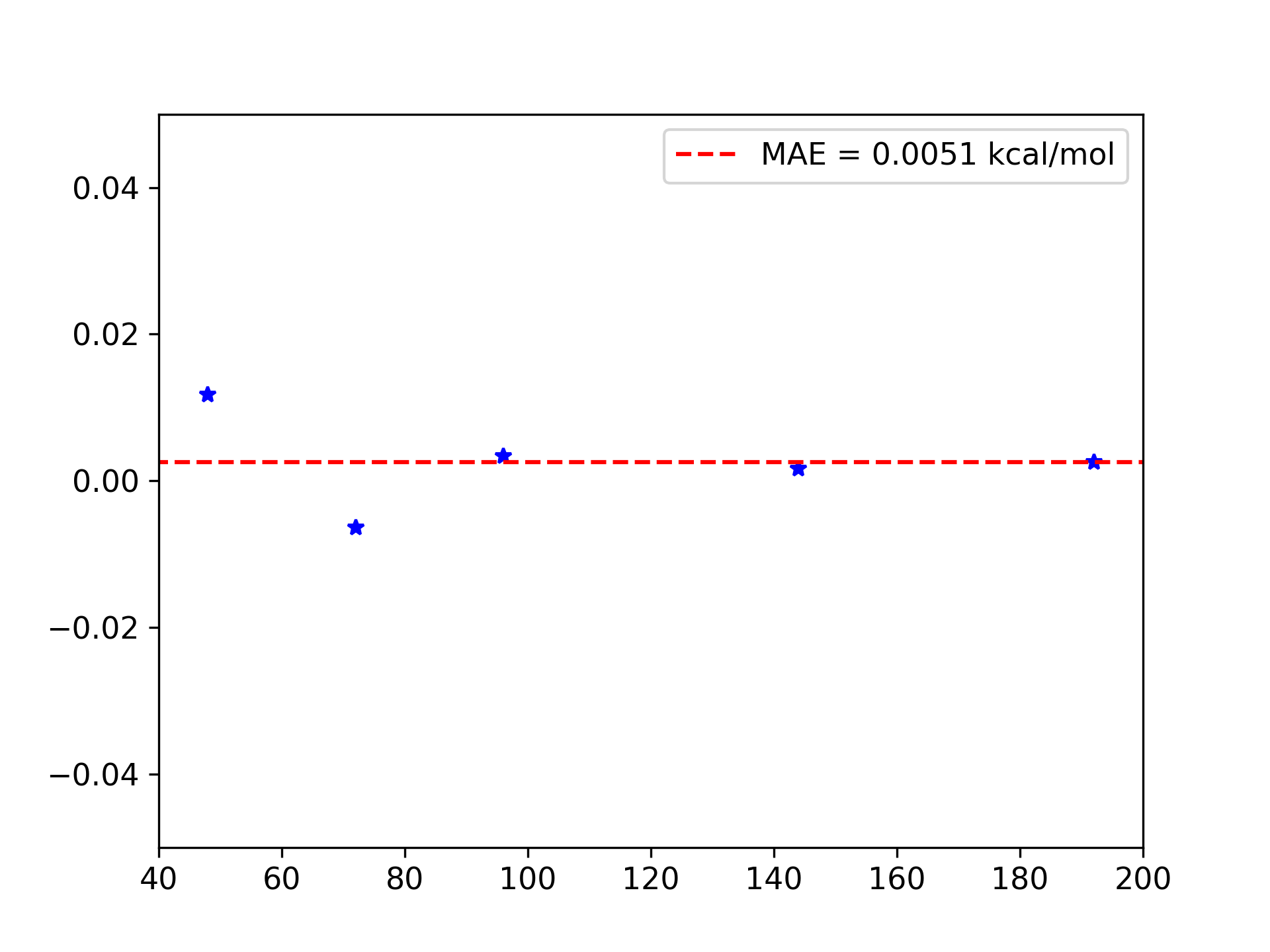}
      }
      \caption{(a,b) The time costs of glycine chains and water clusters by RI-RPA and BTD-RPA (time of each step is shown); (c,d) the error per atom of BTD compared to RI-RPA.}
      \label{gly-rpa}
\end{figure*}
In order to balance accuracy and efficiency, the parameters of the BTD calculation were optimized on the S66 test set of non-covalent interactions.\cite{S66_2011} The reference correlation interaction energies were obtained from RPA@B3LYP/cc-pVDZ calculations. More details of optimization are shown in our previous work.\cite{btd} The number of optimization steps is 120 and there are 20 samples in one step. The test data set is chosen as S66$\times$8 and the references are chosen as B3LYP, BLYP and HF.\cite{S66X8} The training process and test results of BLYP reference are shown in Figure \ref{s66x8}. The test results of B3LYP and HF are collected in Table S3 and Figure S1 of the supporting information

The mean absolute error (MAE) for the non-covalent interactions in S66$\times$8 is 0.08 kcal/mol, and the maximum error is 0.39 kcal/mol. Benchmark calculations on the S66$\times$8 dataset demonstrate the accuracy of BTD-RPA for non-covalent interactions, with errors sufficiently low for precise relative energy determination. The tests of BTD-RPA at the references of HF and B3LYP, which also show the similar bias.

To assess the computational efficiency of the BTD-based algorithm, BTD-RPA calculations are carried out for glycine chains and water clusters, with timings and energies compared to conventional RI-RPA. The geometries of the Water clusters are taken from our previous work,\cite{btd} and those of the glycine chains are taken from literature.\cite{dlpno-rpa} The computational consumption of each step and the total times for the BTD-RPA calculations are shown in Figure \ref{gly-rpa}.

For the test systems ranging in size from 300 to 1600 basis functions, the BTD-RPA method exhibits $O(N^2)$ scaling, which is significantly lower than the $O(N^4)$ scaling of the RI-RPA approach. The time consumption of BTD-RPA is decomposed into three parts: the generation of the BTD (red), calculation of $\bm{\Pi}(i\omega)$ (green), and computation of correlation energy (blue). The most time-consuming step is to generate the BTD interpolative vector with scaling of $O(N^{1.5})$. The calculations of $\bm{\Pi}$ exhibit the cost of $O(N^{2.0})$ and $O(N^{2.3})$ for 1D and 3D systems, respectively. The test results show that the BTD-based algorithm is efficient for both 1D and 3D systems. The deviations of BTD-RPA from RI-RPA are remarkably small, approximately $-0.015$~kcal/mol per atom for glycine chains and $0.05$~kcal/mol per atom for water clusters. This demonstrates that the BTD approximation introduces negligible error in energy evaluations while offering superior computational efficiency.

\begin{figure}
    \centering
    \includegraphics[width=1.0\linewidth]{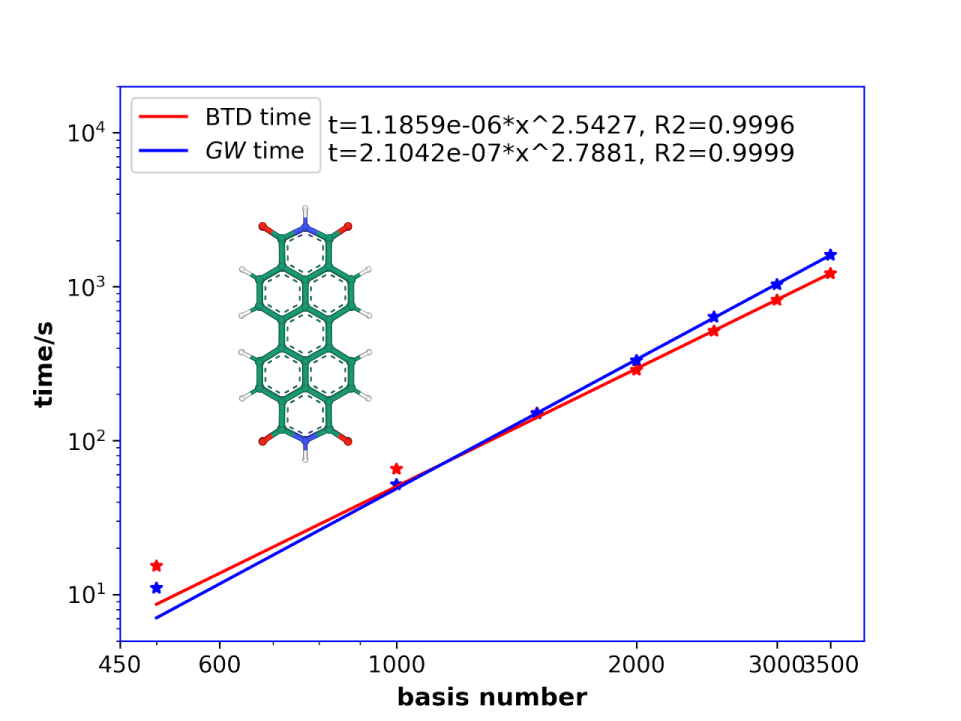}
      \caption{Scaling of BTD-ev$GW$ computational time with system size for $\pi$-stacked aggregates (1–7 monomers) with the cc-pVDZ basis set}
      \label{evGW}
\end{figure}

The scalability of BTD-ev$GW$ algorithm is demonstrated by a series of $\pi$-conjugated systems in an H-aggregation (face-to-face) stack, as illustrated in Figure \ref{evGW}. The systems range from a single monomer to a heptamer, with the largest system comprising 280 atoms and 3500 basis functions. The values of HOMO and LUMO are shown in Figure S2. The scaling analysis presented in Figure \ref{evGW} reveals the favorable power-law exponents: approximately $O(N^{2.5})$ for the BTD kernel generation and $O(N^{2.8})$ for the self-energy and dynamic screening steps. A BTD-ev$GW$ calculation for the largest system (3500 basis functions) converges in five self-consistent cycles and completes in approximately 6000 seconds. This robust performance confirms that our BTD-$GW$ formalism is both efficient and applicable to large-scale molecular systems.

\section{Conclusion}
In summary, we have developed a formally cubic-scaling $GW$ algorithm by integrating the BTD framework with Laplace-transform-based imaginary-time/frequency conversions  and analytic continuation. By exploring the sparsity of the density-density response in real space, the time scale can be further reduced to about $O(N^2)$ for the calculation of $\bm{\Pi}$. The parameters for BTD are optimized on the S66 data set to balance both accuracy and efficiency. The benchmark results confirm that BTD-RPA delivers excellent accuracy for relative energies while achieving sub-quadratic scaling ($< O(N^2)$) for both 1D and 3D systems ranging from 300 to 1600 basis functions. Crucially, our BTD-evGW approximation enables GW calculations for systems with over 3000 basis functions, bringing the practical application of this accurate method to large molecular systems. Future work will also assess the accuracy of BTD-$GW$ for systems containing heavier elements and larger basis sets, further establishing its robustness across vari chemical environments.

The BTD-$GW$ quasiparticle energies provide a natural starting point for subsequent Bethe–Salpeter equation (BSE) calculations (BSE@$GW$), enabling accurate prediction of molecular excitation spectra. When combined with the BTD algorithm for computing the exchange kernel, the BSE@$GW$ approach can also achieve formal $O(N^3)$ scaling, paving the way for large-scale excited-state calculations. The related work is going on. The calculation of $GW$ still suffers from the long-range performance of the Coulomb interaction. How to screen long-range interaction for self-energy's calculation is still a problem.

\section*{Supplementary Material}
    See supplementary material associated with this article, including Tables S1-S3 and Figures S1-S2 and and the zip files of the coordinates for the geometries of aggregations from 1 to 7.

\begin{acknowledgments}
    This project is supported by the National Natural Science Foundation of China (Nos. 22473092, 22173076, 22373077).
    \end{acknowledgments}
    
    \section*{AUTHOR DECLARATIONS}
    \subsection*{Conflict of Interest}
    The authors have no conflict of interest to disclose.
    \section*{Data Availability Statement}
The data that support the findings of this study are available
within the article and its supplementary material.

    \section*{reference}
    \bibliography{aipsamp}
\end{document}